\documentclass[preprint]{elsarticle}

\usepackage{graphicx,amssymb,url}
\usepackage{wasysym}
\usepackage{dcolumn}% Align table columns on decimal point
\usepackage{bm}% bold math
\usepackage{color}
\usepackage{array}

\begin{document}

\begin{frontmatter}

\title{Modeling the clustering in citation networks}

\author[ict]{Fu-Xin Ren\corref{cor}}
\ead{renfuxin@software.ict.ac.cn}
\author[ict]{Hua-Wei Shen}
%\ead{shenhuawei@software.ict.ac.cn}
\author[ict]{Xue-Qi Cheng}
%\ead{cxq@ict.ac.cn}

\address[ict]{Institute of Computing
Technology, Chinese Academy of Sciences, Beijing 100190, PR China}
\cortext[cor]{Corresponding author}

% \pacs{89.75.Hc}{Networks and genealogical trees}
% \pacs{89.75.Fb}{Structures and organization in complex systems}
% \pacs{89.20.Ff}{Computer science and technology}

\begin{abstract}
For the study of citation networks, a challenging problem is modeling
the high clustering. Existing studies indicate that the promising way
to model the high clustering is a copying strategy, i.e., a paper
copies the references of its neighbour as its own references. However, the
line of models highly underestimates the number of abundant triangles
observed in real citation networks and thus cannot well model the high
clustering. In this paper, we point out that the failure of existing
models lies in that they do not capture the connecting patterns among
existing papers. By leveraging the knowledge indicated by such connecting
patterns, we further propose a new model for the high clustering in
citation networks. Experiments on two real world citation networks,
respectively from a special research area and a multidisciplinary
research area, demonstrate that our model can reproduce not only the
power-law degree distribution as traditional models but also the
number of triangles, the high clustering coefficient and the
size distribution of co-citation clusters as observed
in these real networks.
\end{abstract}

\end{frontmatter}

\section{Introduction}
As a concise mathematical tool, network is widely used to describe
the systems of interacting components \cite{Watts1998,Barabasi1998,Newman2003},
including social networks, World Wide Web and citation networks,
to name a few. Among the studies on networks, much research
attention has been paid to citation networks of papers, patents
and legal cases \cite{Price1965,Redner1998,Lehmann2003,Zhu2003,Redner2005}.
In particular, the scientific citation networks are the research subjects 
of much literature and it is believed that such studies can help us 
better understand the collaboration of scientists, the exchange of 
ideas and create better scientific impact measures. In this paper, 
we will focus on scientific citation networks. 

One outstanding challenge of the studies on citation networks is
to find the mechanism which governs the growth of citation networks.
For this purpose, many works have been done to investigate and model
the growth of citation networks \cite{Sen2005,Hajra2005,Hajra2006,Cheng2007,Wang2008,Wang2009,Eom2011}.
Among the methods for citation network modeling, growth models are
widely used with the considerations that papers in citation network
are added sequentially and all the out-links of a paper are generated
when it joins the network. In a growth model, the key is to determine
the papers which will be cited by the new paper. Existing models
addressed such problem using certain preferential attachment mechanisms,
involving the in-degree \cite{Price1965,Redner2005,Jeong2003},
the age \cite{Zhu2003,Hajra2005,Hajra2006,Wang2008,Wang2009,Dorogovtsev2000}
and the content similarity \cite{Menczer2004,Cheng2009}.
These models perform well at reproducing the power-law degree distribution.
However, they underestimate the number of triangles and thus
fail to model the high clustering in citation networks,
which is closely related with network transitivity and
the formation of communities \cite{Kumpula2007}.

\begin{figure}
\begin{center}
\includegraphics[width=8cm]{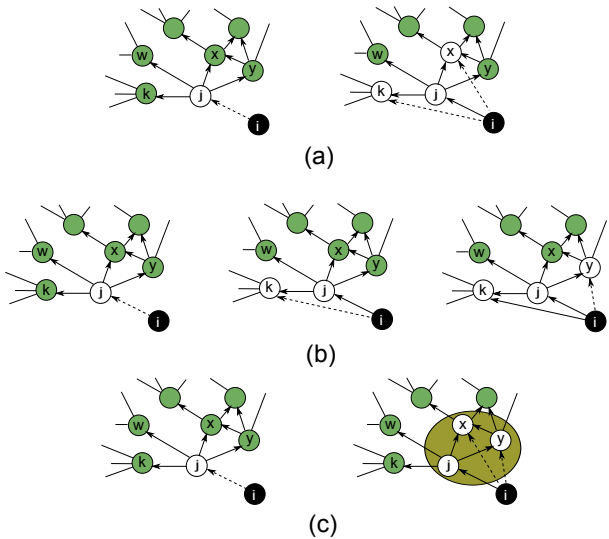}
\caption{\label{fig1} Illustrations of (a) the forest fire model, (b) the
triadic closure model and (c) the DAC model proposed in this paper.
Here the node $i$ is a new node and we assume its out degree is $3$.
In the forest fire model, $i$ firstly connects to an old node $j$ randomly
and then links to some of $j$'s out- and in- neighbours, $k$ and $x$ for example.
In the triadic closure model, $i$ firstly connects to an old node $j$ through preferential attachment
and then with some probability links to one of $j$'s neighbours, node $k$ as an example.
Then $i$ attach an arc to one of $j$'s or $k$'s neighbours, $y$.
In the DAC model, $i$ firstly connects to an old node $j$ according to
preferential attachment and then connects to $j$'s neighbours considering
the connecting pattern among them, such as the clique structure. Here,
nodes $j$, $x$, $y$ form a clique and thus are preferred to be connected by node $i$.
}
\end{center}
\end{figure}

\begin{figure}
\begin{center}
\includegraphics[width=6cm]{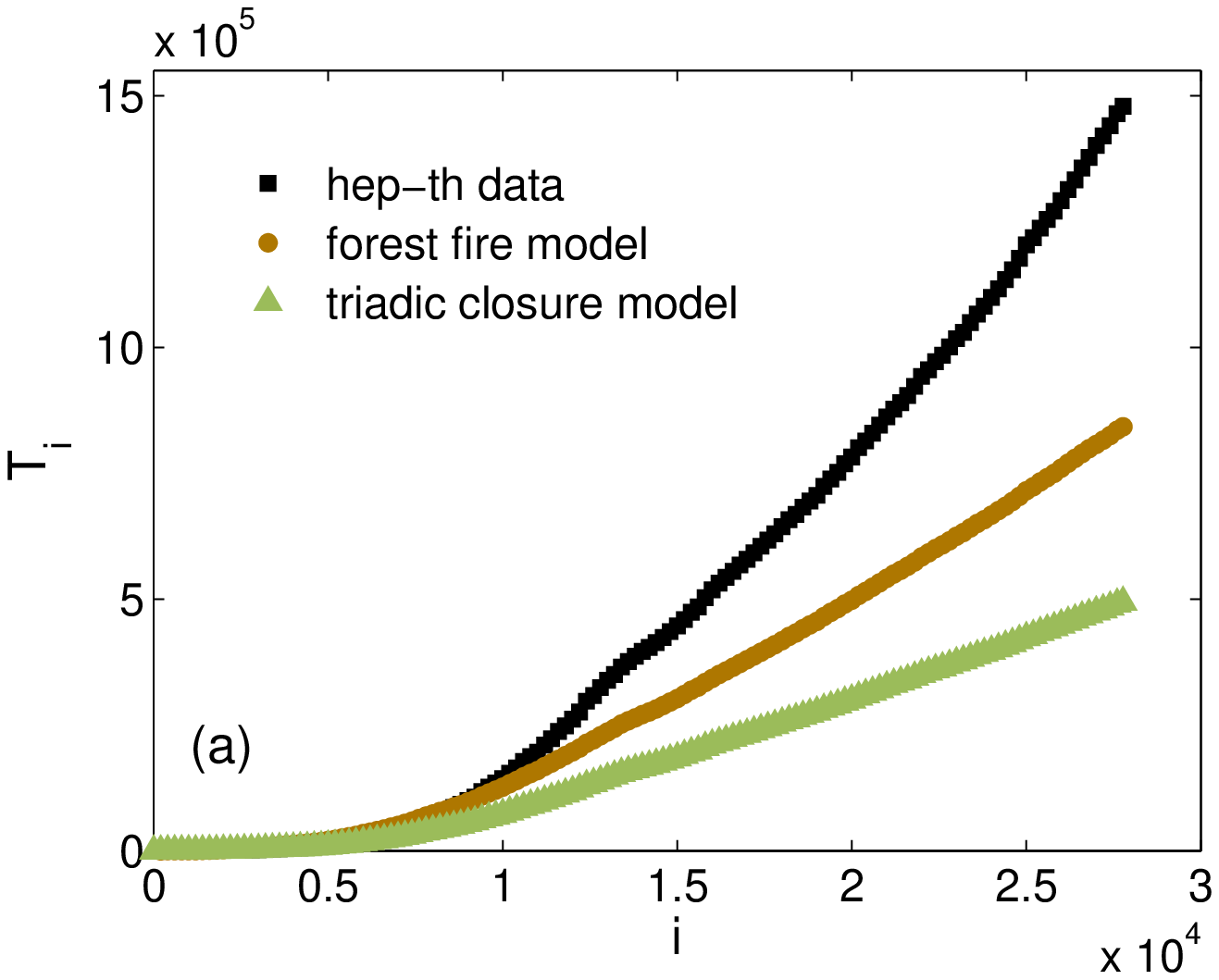}
\includegraphics[width=6cm]{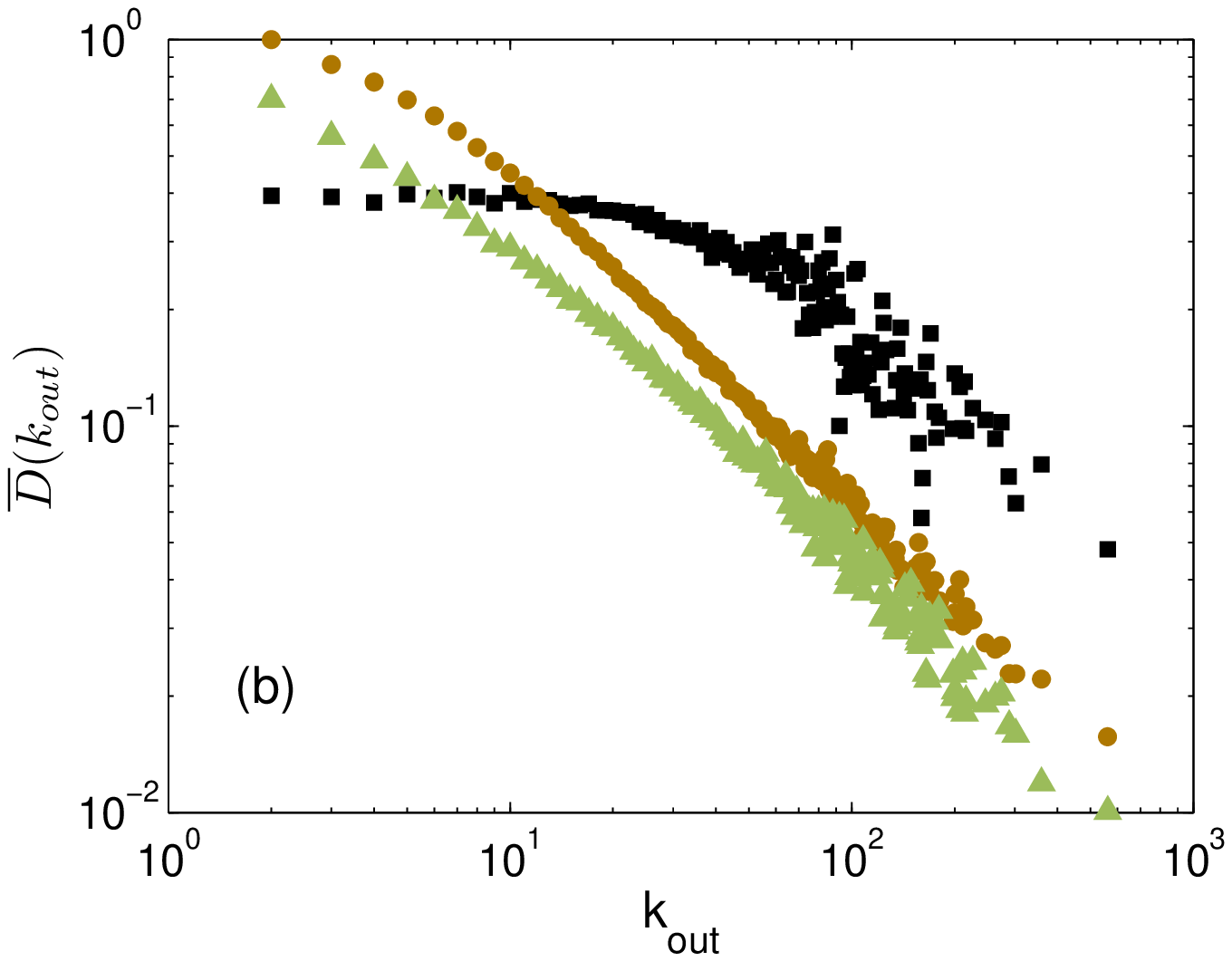}
\caption{\label{fig2} (a) The growth of triangle number
$T_i$ as a function of the network size $i$ of hep-th data, forest fire model
and triadic closure model for the data. The
hep-th data depict the citation relations among the preprints
on high-energy theory archive posted at www.arxiv.org between 1992
and 2003. For the forest fire model the parameters are the
same as in \cite{Leskovec2005}, and for the triadic closure model
the parameters are the same as in \cite{Wu2009}. (b) The link density of
the reference graph as a function of node's out-degree in the hep-th data, forest fire model
and triadic closure model for the data. $k_{out}$ denotes
the out-degree of the node and $\overline{D}$ is the average link density of
reference graphs of nodes with the same out-degree.
}
\end{center}
\end{figure}

The common practice to produce triangles is a copying strategy \cite{Kumpula2007,Holme2002},
i.e., a node copy the links of its neighbour as its own, 
partially or completely. Two typical models are the
forest fire model proposed by Leskovec \textit{et al.} \cite{Leskovec2005} and the
triadic closure model proposed by Wu \textit{et al.} \cite{Wu2009},
as shown in Fig.~\ref{fig1}. In the forest fire model, a new paper
randomly cites an existing paper and then cites its references
and its citing papers with certain probability. In the triadic
closure model, a new paper either cites an existing paper
according to certain preferential attachment mechanism or cites
the papers cited by the new paper's references. To our surprise,
although these two typical models are designed with the goal to
form abundant triangles, they highly underestimate the number of
triangles observed in real world networks, as shown in Fig.~\ref{fig2}(a)\footnote{
In \cite{Wu2009} the number
of triangles is claimed to agree with the real data. However,
lots of the generated triangles are duplicate and in this paper the
results are calculated after removing those duplicates.}.
One possible cause of the underestimation lies in the copying strategy
to form triangles. Specifically, when a new paper copies the links
of its neighbours, it ignores the existing connections among the targets
which are the papers citing or cited by the new paper's references.
As shown in Fig.~\ref{fig1}, both the forest fire model and the triadic
closure based model are blind to the fact that there exits an link between
the target papers $x$ and $y$ and thus miss the chance to form more
triangles through citing them.

In this paper, by leveraging the knowledge ignored by the forementioned
two models, we propose a new model to model the high clustering in
citation networks. We further verify the effectiveness of our model
using two real world citation networks, respectively from a
special research area and a multidisciplinary research area.
Experimental results demonstrate that our model can reproduce not
only the power-law degree distribution as traditional models
but also number of triangle, the high clustering coefficient and the
size distribution of co-citation clusters
as observed in these real networks.

The rest of this paper is organized as follows. In Section~\ref{sec2},
we analyze the structural characteristics of the reference graph of
papers in a real citation network. Here, reference graph of a paper
characterizes citation relations among the references of this paper.
Based on the analysis results, in Section~\ref{sec3}, we propose
our DAC model to modeling the high clustering in citation networks.
Section~\ref{sec4} describes the experimental results by applying
our model to model two real networks. Finally, Section~\ref{sec5}
concludes this paper and gives some discussions.

\section{The reference graphs in the real data}
\label{sec2}

Before giving a model for citation network, we first analyze
a real world citation network, the hep-th network, 
to provide some intuitive indications for designing an appropriate 
model. Our analysis is conducted on the reference graph of
each paper. A \emph{reference graph} of a paper characterizes the
citation relations among the references of the paper. For a given paper,
its reference graph can be viewed as its ``ego-graph'' or ``ego-network''
but excluding itself and the papers citing it. The structure of a reference
graph provides us a complete picture about the connecting status among
papers before they are really cited. Therefore, the analysis on such a
graph is critical to find clues for the microscopic mechanisms 
governing the evolution of citation networks.

\begin{figure}
\begin{center}
\includegraphics[width=6cm]{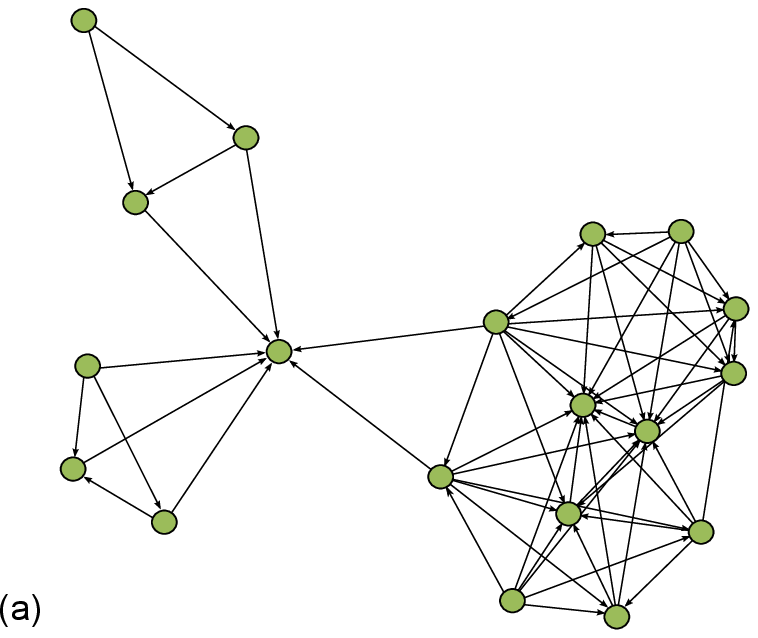}
\includegraphics[width=6cm]{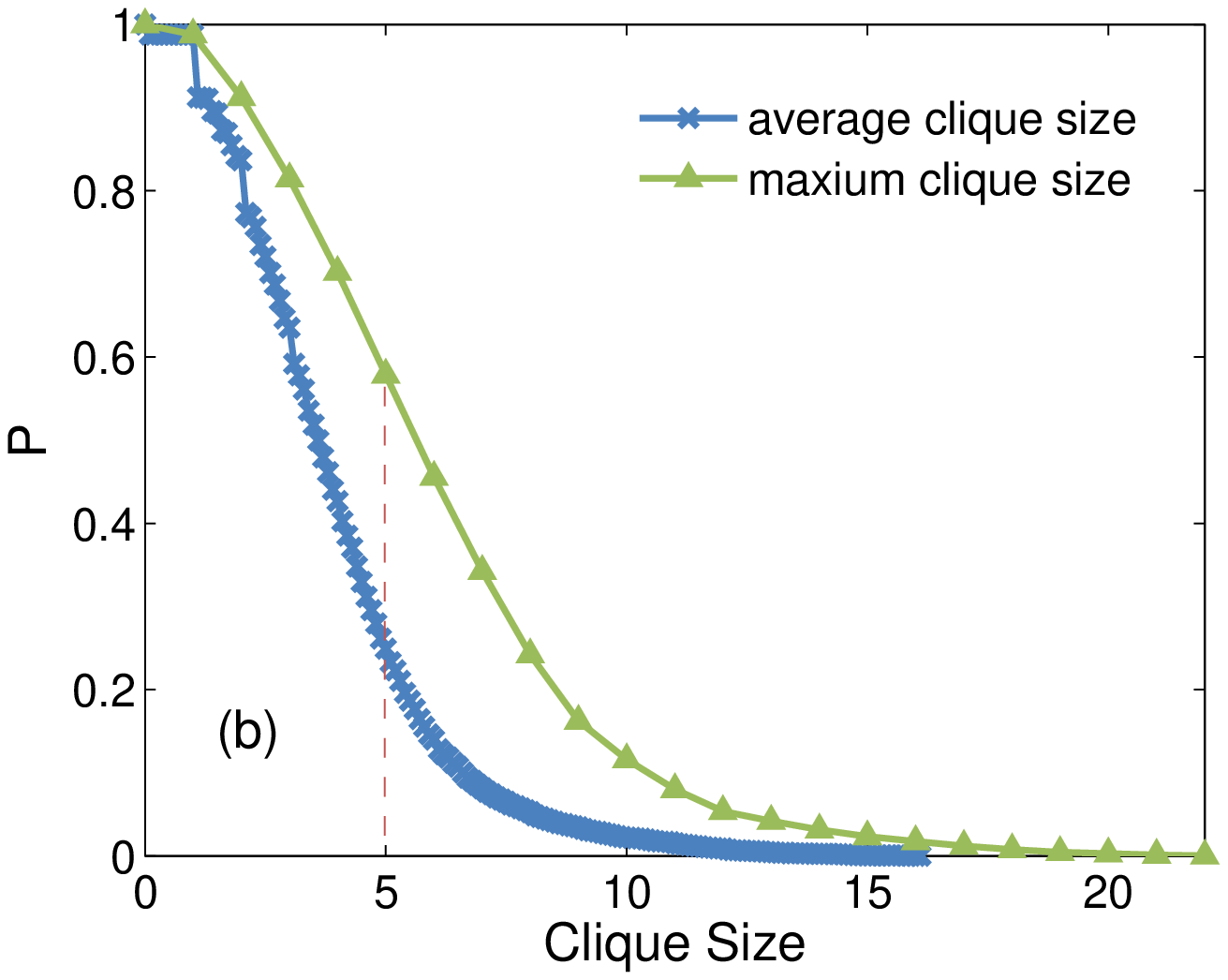}
\caption{\label{fig3} (a) The reference graph
of node $19000$ in the hep-th data. This node corresponds to
the paper ``Brane world from IIB matrices''
with DOI 10.1103/PhysRevLett.85.4664. The graph contains $19$ nodes
and $58$ arcs. (b) The cumulative distributions $P$ of
the average clique size and maximum clique size in the reference graphs.
For the average clique size of one reference graph, we find out
all the maximal cliques in the reference graph, sum up their sizes and then
make an average. While for the maximum clique size, it refers to the size
of the largest maximal clique in the reference graph. We can
see that about 58\% reference graphs have a maximum clique with size
no less than 5, and about 25\% graphs have the average clique
size larger than 5. Moreover, 11\% graphs have a maximum clique
no less than 10. Old nodes have few out-degrees
and to make observations clear here the result is gained
using the reference graphs of the newest 20\% nodes.
}
\end{center}
\end{figure}

As an example, Fig.~\ref{fig3}(a) shows the reference graph of a paper
in the hep-th data.
We can see that nodes in the graph are connected into a single component.
This indicates that when authors cite one paper they also tend to
cite the paper's neighbours, i.e.,
papers in the paper's references or papers citing the paper.
This phenomenon reflects the reading behaviour of researchers, i.e.,
when they are interested in a paper they are very likely to be interested
the papers in its references and papers citing it. From Fig.~\ref{fig3}(a),
we can also find that the reference graph has a very high link density.
Fig.~\ref{fig2}(b) shows the link density of reference graphs with respect
to the out degrees of papers. It is clear that the link density of a paper's
reference graph is correlated with its out-degree. This phenomenon may be
attributed to the facts that papers with high out-degrees are usually reviews
or surveys and thus their reference graph have lower link density while
papers with low out-degrees are papers on a specific topic. Furthermore,
we can find such a phenomenon cannot be well modeled by the existing two
typical models for high clustering in citation network. In particular,
the link density of the papers with high out-degrees are largely
underestimated.

We further find that the reference graph contains many cliques
with large sizes. A clique is a subgraph within which every two nodes
are connected. Abundant cliques are crucial to high clustering \cite{Cheng20102} and
community structure \cite{Palla2005,Shen2009,Shen20092}.
As shown in Fig.~\ref{fig3}(a), the reference graph contains
two cliques with size $7$ and the
average clique size of the graph is about $4.56$. A large clique may
contain many smaller cliques. In this paper, we use the maximal
clique to avoid the repetitive counting. Fig.~\ref{fig3}(b) illustrates
the distribution of the size of maximal cliques.
The formation of these cliques roots in that authors
always cite a group of papers which are closely related.
Take the literature of research on citation network as an
example:
In 2005, a paper $k$
\cite{Redner2005} revealed long-term systematic features of citation
statistics based the observations on a period of real data.
Later on, a paper $j$ \cite{Hajra2006} provided
a model for the aging characteristics in citation networks and
cited $k$ as a reference. Recently, Wu \emph{et al.}'s paper $i$ \cite{Wu2009}
integrated the aging and triadic closure mechanisms to model
the citation patterns and cited both $j$ and $k$,
which brings a 3-clique $ijk$.
As research on this problem goes on, new papers (such as this paper)
will cite these formers and larger cliques will emerge.
Thus, highly connected structure, such as clique, indicates topical
correlations among the nodes in it. When a paper cites one node in a
clique, with a high probability
it will cite others also in the clique.
Besides, a paper prefers to cite those with large in-degree (popular) and
small age (undergoing recognition). Therefore, in a growth model
in-degree and age are always taken into the preferential attachment.

\section{The DAC model}
\label{sec3}

On the basis of above observations, we propose our model
for citation networks - the \emph{Degree}-\emph{Aging} preferential attachment
and \emph{Clique} neighbourhood attachment model, DAC model for short.
It is a growth model in which nodes join the network sequentially and attach their
arcs to the old ones. In citation networks, nodes are ordered temporally, i.e., they joined
the network according to their ages. In our DAC model we
keep the orders and out-degree of nodes the same as in the original data.
It is innocuous to take the out-degree as given information because the out-degree
of each paper is decided by its authors and most of the time we concern about the in-degree.
As its name explains, the DAC model is composed of two parts,
\begin{itemize}
  \item \emph{the degree-aging preferential attachment}.
  A new node $i$ firstly originates an arc to an old
  node $j$ according to the probability $\prod_{ij} \propto k_{j}^{in} \times t_{j}^{-\alpha}$,
  where $k_{j}^{in}$ is $j$'s in-degree, $t_{j}=i-j$ is the age of $j$
  and $\alpha>0$ is the decaying parameter. 
  Actually, this power-law form of probability function is widely 
  adopted to model degree-aging preferential attachment 
  in the literature, such as done in the Dorogovtsev-Mendes (DM) 
  model \cite{Dorogovtsev2000} and the model in \cite{Hajra2006}.

  \item \emph{the clique neighbourhood attachment}.
  With probability $\beta$ ($0\leq\beta\leq1$), node $i$ chooses to
  link $j$'s \emph{clique neighbours}, i.e., the nodes in the same
  clique $j$ belongs to, as illustrated in Fig.~\ref{fig1}(c).
  Node $j$ may belong to many cliques
  and $i$ randomly chooses one proportional to the clique's size and links
  all the nodes in the clique.
  Otherwise, i.e., with probability $1-\beta$ or when there are no
  clique neighbours $i$ can connect to, $i$ attaches an
  arc using the degree-aging preferential attachment
  as above. Here $j$ is one of $i$'s neighbours.
\end{itemize}

We repeat above attachment mechanisms to fill up the remaining out-degrees of $i$.
Obviously, the clique neighbourhood attachment takes the connecting patterns of
the potential neighbours into account and guides the formation of triangles.
By tuning the parameter $\beta$ we can control the growth rate
of clustering, i.e., larger $\beta$ produces larger clustering.

\section{The data and modeling results}
\label{sec4}

In this section, we examine the DAC model on the following two real-world citation networks.
\begin{itemize}
\item \texttt{hep-th} data, which comes from preprints on the high-energy theory
archive posted at www.arxiv.org between 1992 and 2003. It contains $27,770$
preprints after cleaning.

\item \texttt{PNAS} data, which contains $23,572$
articles published by the Proceedings of the National Academy of
Sciences (PNAS) of the United States of America from $1998$ to
$2007$. We crawled the data at the journal's website
(\url{http://www.pnas.org}) in May 2008\footnote{We removed the
isolated nodes in the two data as we are going to model the citation patterns
of citation networks and these nodes matter nothing in this study.}.
\end{itemize}

We choose the two networks because they provide data with
different types, i.e., one is on a special research area
and the other is on multidisciplinary sciences.
The basic structural statistics of the two data are listed in
table~\ref{tab:Data_description}. It shows that the two networks
are comparable in network size while the hep-th network
is much denser than PNAS. Since a large fraction
of articles on the high-energy theory is put at www.arxiv.org,
the inner citations in the hep-th data is very dense. While for PNAS
data, papers broadly span physical, biological and social sciences,
therefore the inner citations are much lower.

As we intend to model the clustering features in citation networks,
three quantities are observed here: the number of triangles,
the clustering coefficient and the link density of reference graph.
The triangle number of the network is the basic statistic of
clustering structures and its growth as a function of network size
provides insights of how the clustering evolves. The average clustering
coefficient for the network gives an overall indication of the
clustering in the whole network. We also analyze the average clustering
coefficient of vertices with the same degree as a function of the degree,
because this correlation is a useful function to understand the local
structure of the network. For the link density of reference graph,
it is used to validate the matching of the real data and our model
in selecting neighbours. Besides these statistics, the basic structural
quantity, i.e., the in-degree distribution, is also measured here.

\begin{table}[ht]
\caption{
Basic statistics of the hep-th data and PNAS data. $N$,
$L$, $\triangle$ and $\overline{C}$ denote the number of nodes, arcs, triangles and
average clustering coefficient \cite{Watts1998} in the empirical
networks. $\triangle_{ER}$ denotes the triangle number in the networks generated
by the E-R random graph model. $\triangle_{DAC}$ and $\overline{C}_{DAC}$ denote
the triangle number and average clustering coefficient in the networks generated
by the DAC model. The results of E-R model and DAC model are averaged over 100 independent realizations. }
\label{tab:Data_description}
\begin{center}
\begin{tabular}[]{ccc}
\hline \hline
Measures/Networks & hep-th   & PNAS \\
\hline
$N$   & 27,770 & 23,572 \\
$L$ & 352,768 & 40,853 \\
$\triangle$ & 1478,735 & 13,225 \\
$ {\triangle}_{DAC} $ & 1484,004$_{\pm3813}$ & 13,336$_{\pm172}$  \\
$\triangle_{ER}$ & 2742$_{\pm51}$ & 7$_{\pm2}$  \\
$\overline{C}$ &  0.312 & 0.171 \\
$\overline{C}_{DAC}$  & 0.354$_{\pm0.005}$ & 0.186$_{\pm0.002}$ \\

\hline \hline
\end{tabular}
\end{center}
\end{table}

\begin{figure}
\centering
\includegraphics[width=5.5cm]{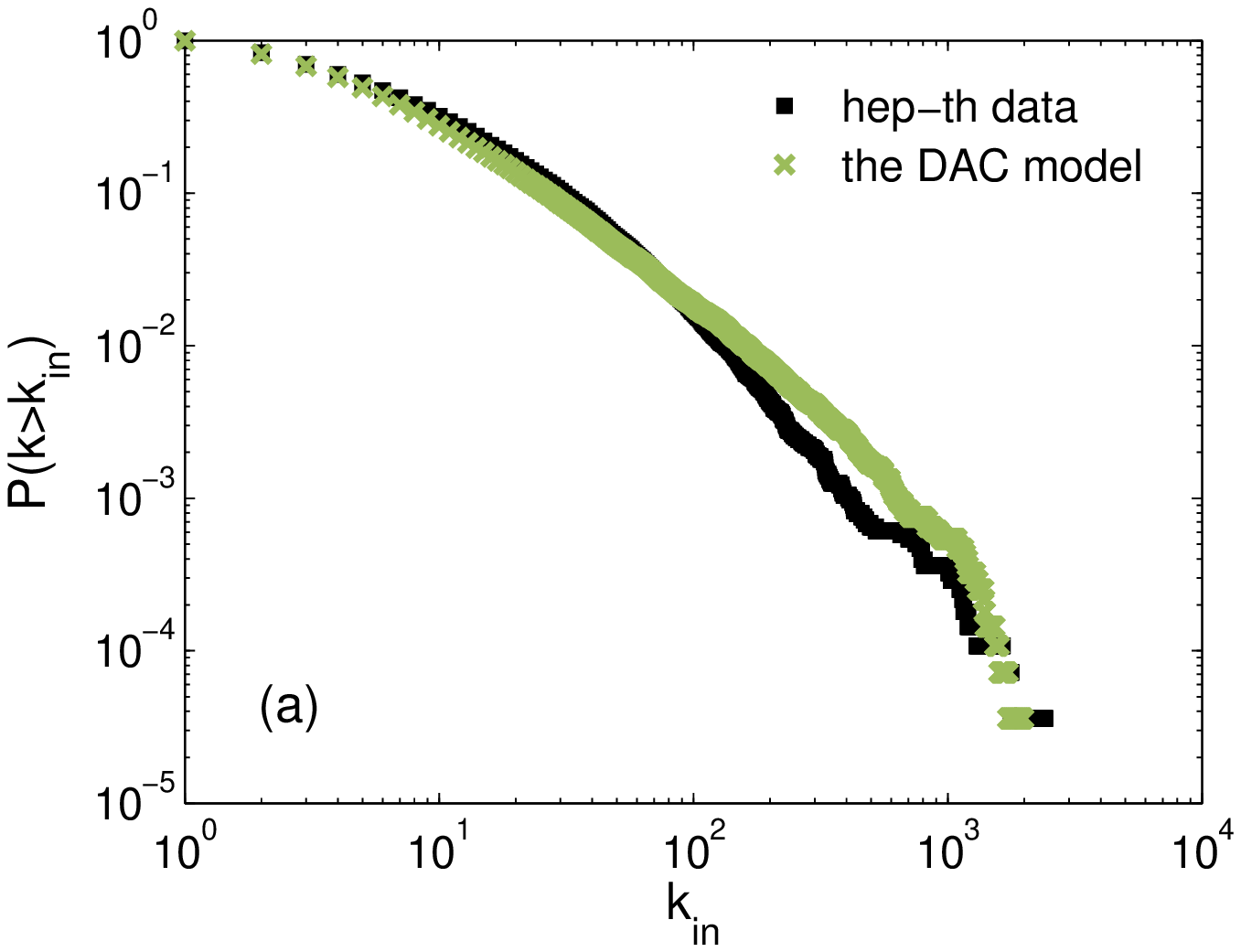}
\includegraphics[width=5.5cm]{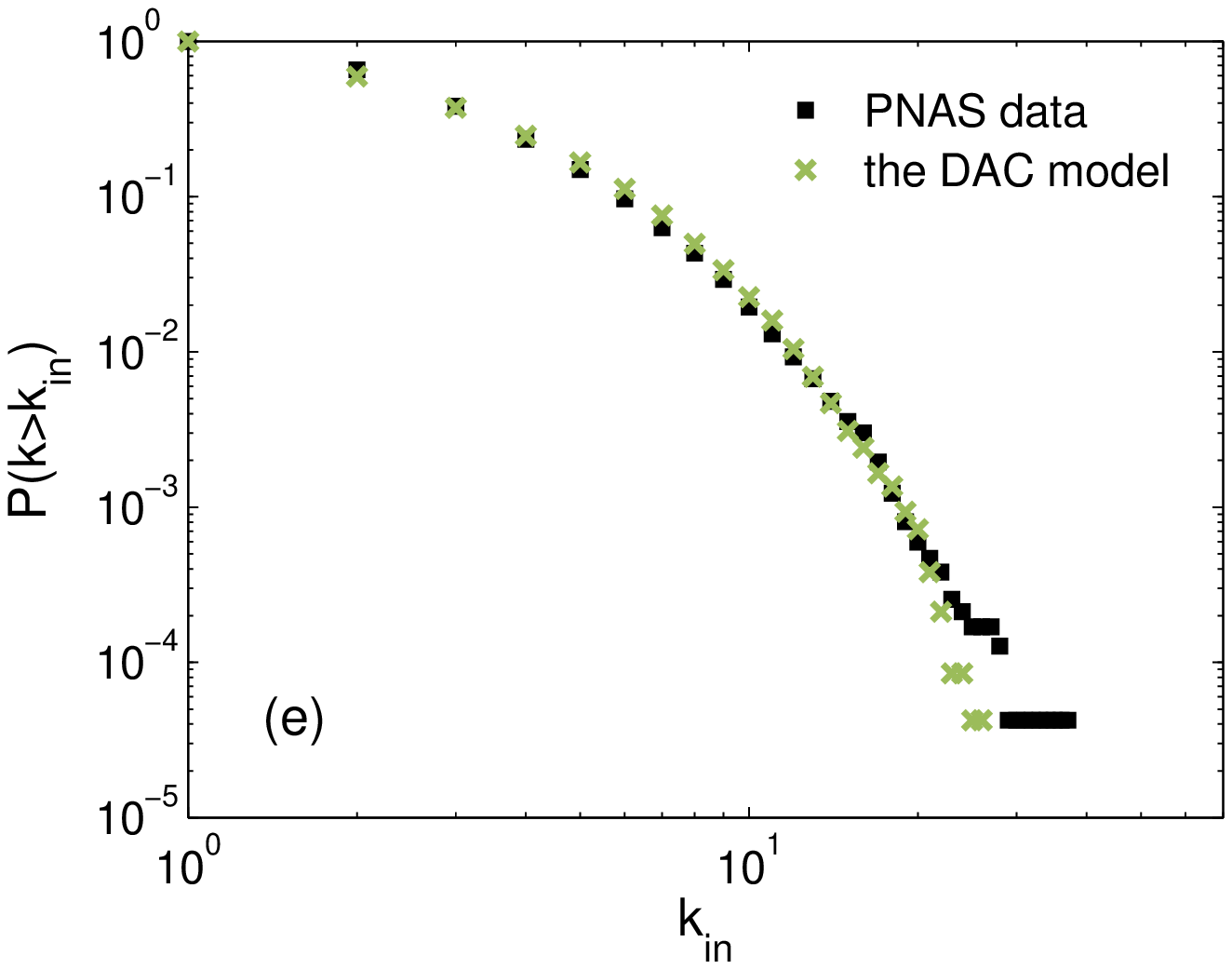}
\includegraphics[width=5.5cm]{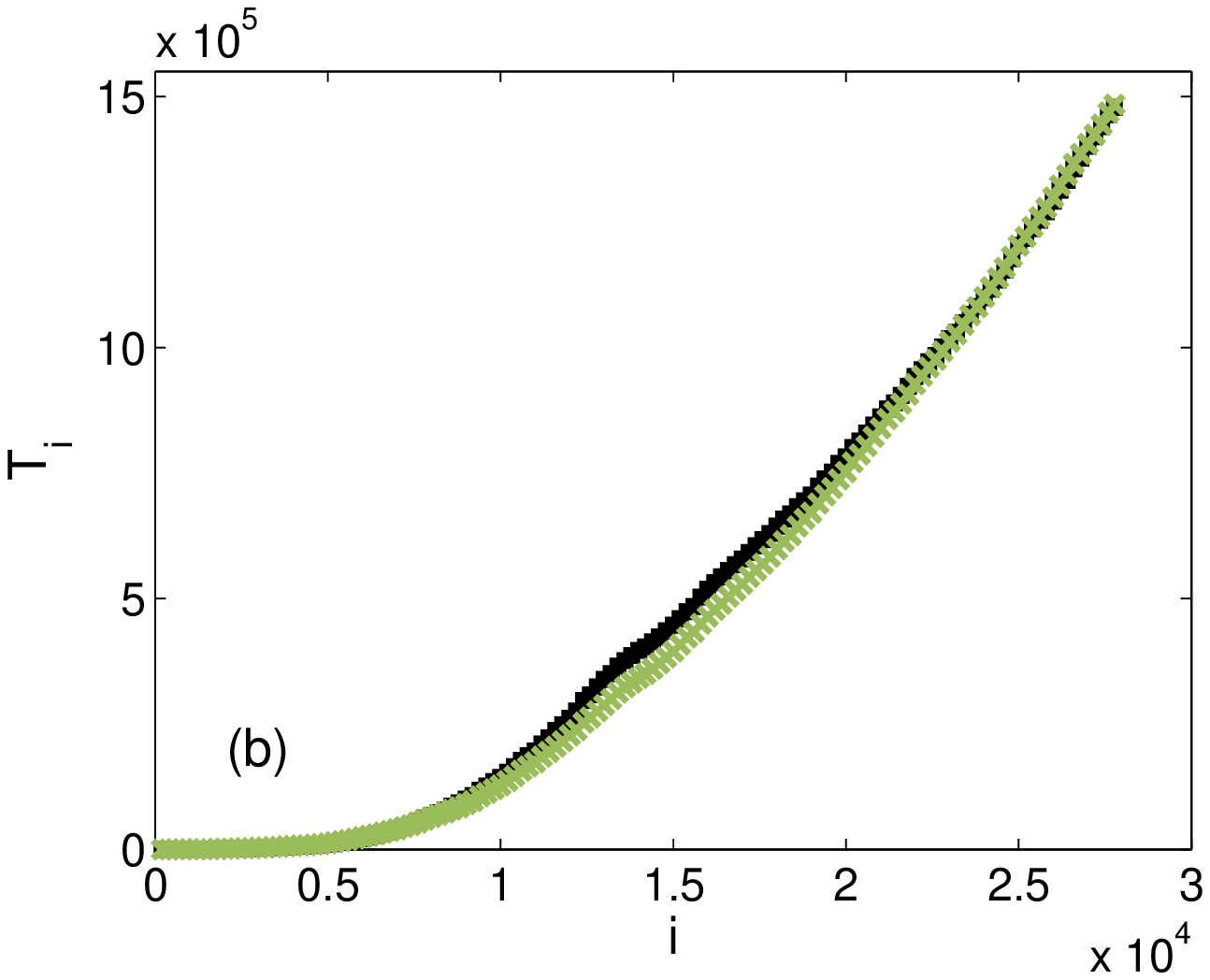}
\includegraphics[width=5.5cm]{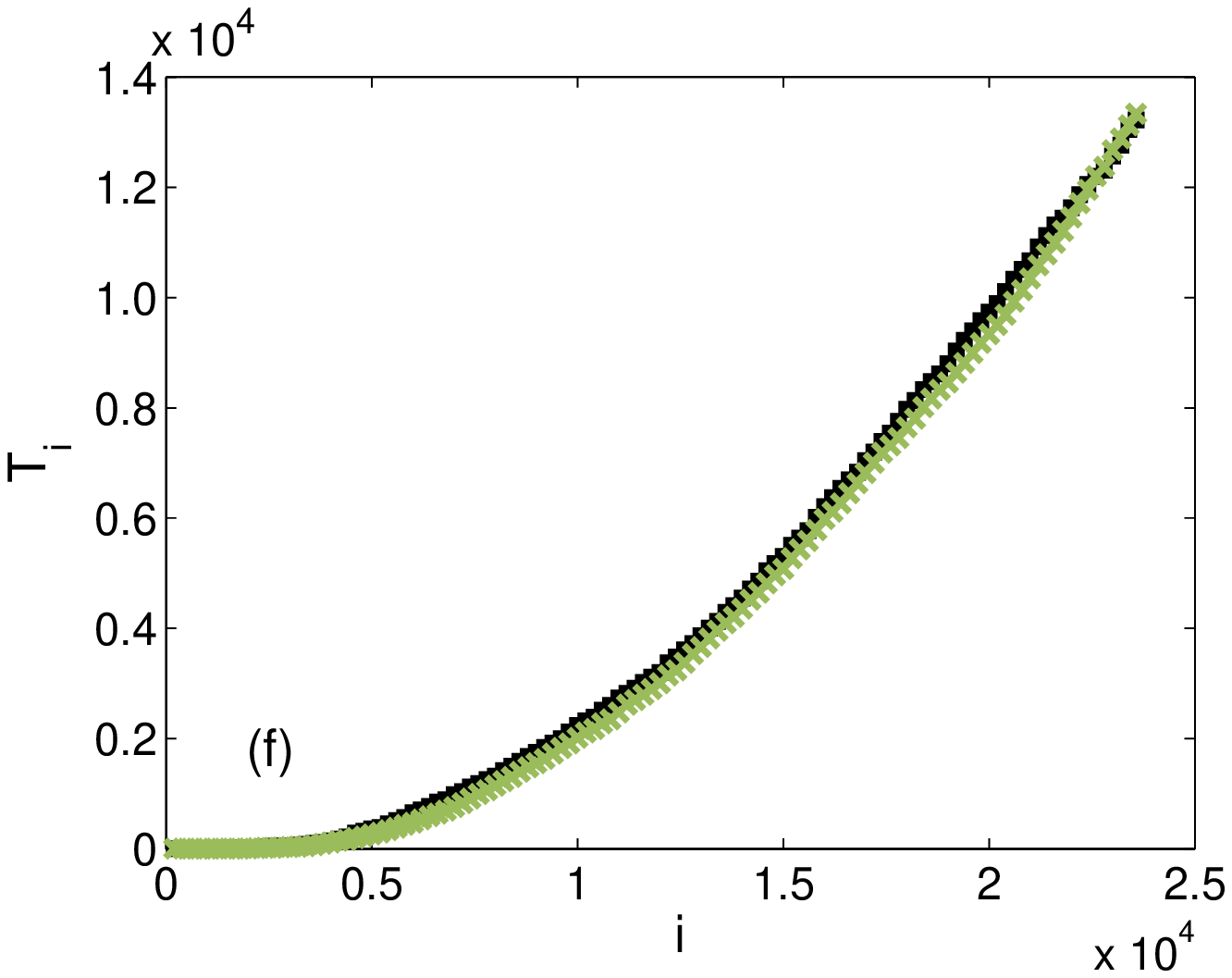}
\includegraphics[width=5.5cm]{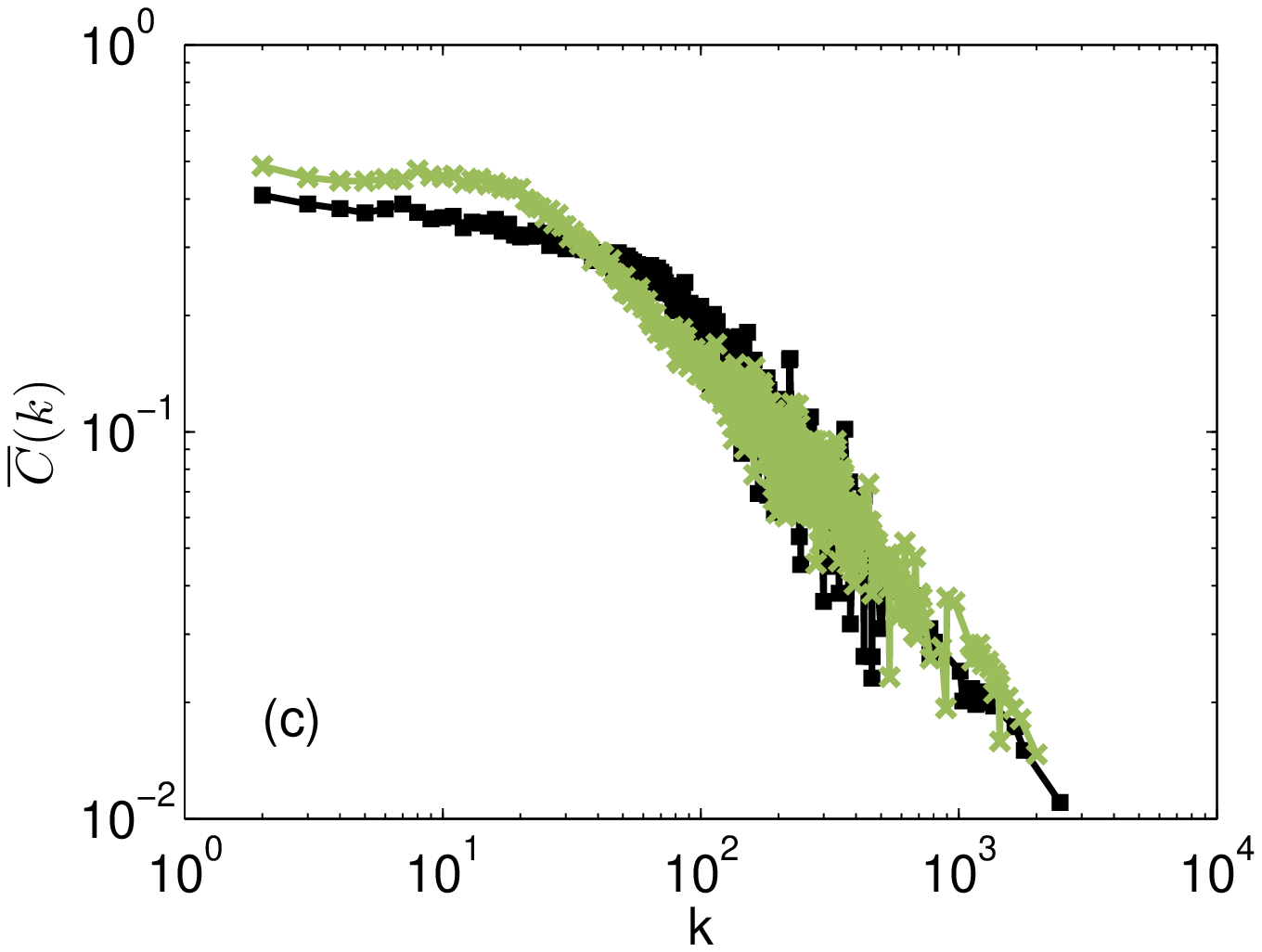}
\includegraphics[width=5.5cm]{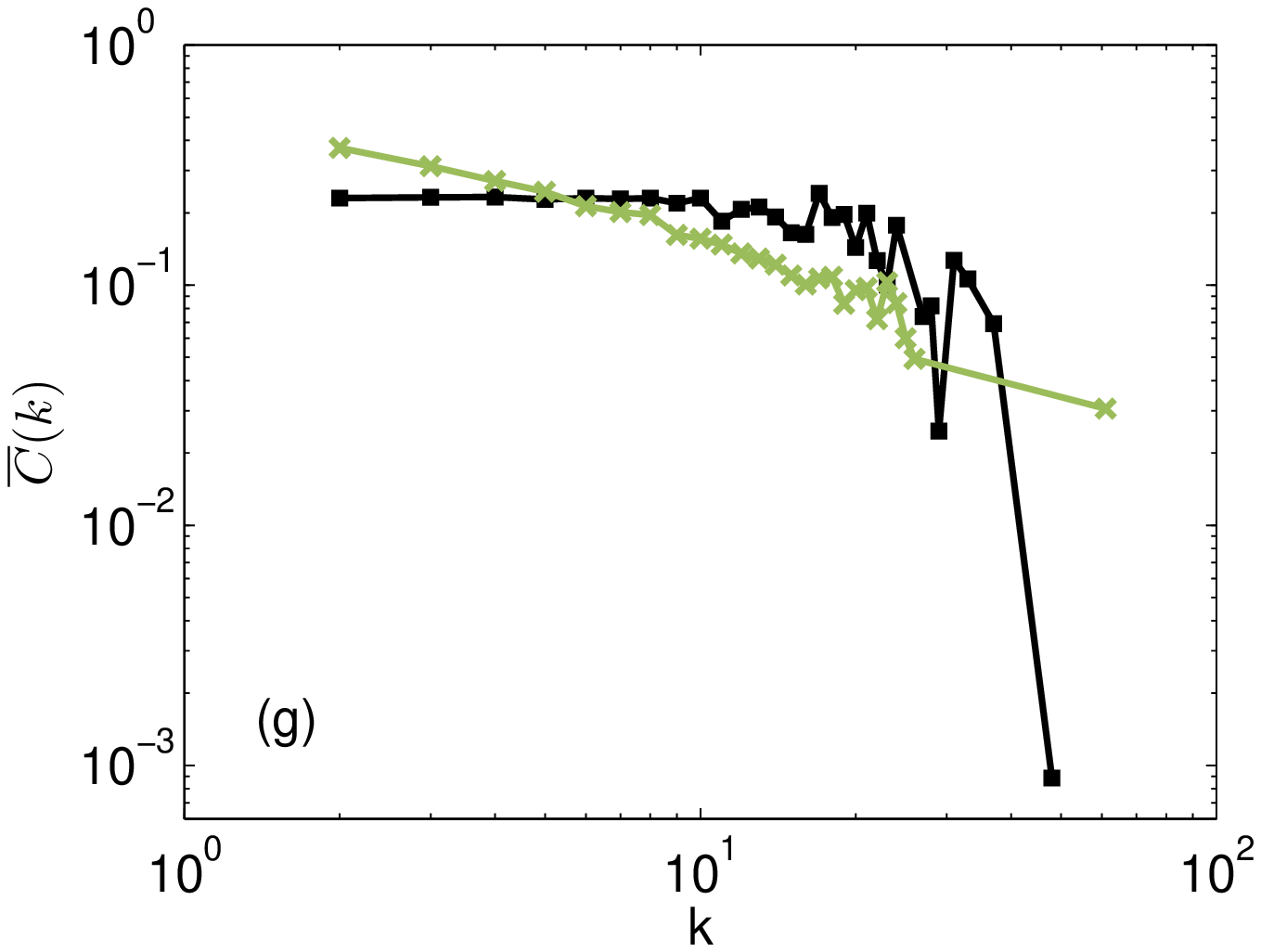}
\includegraphics[width=5.5cm]{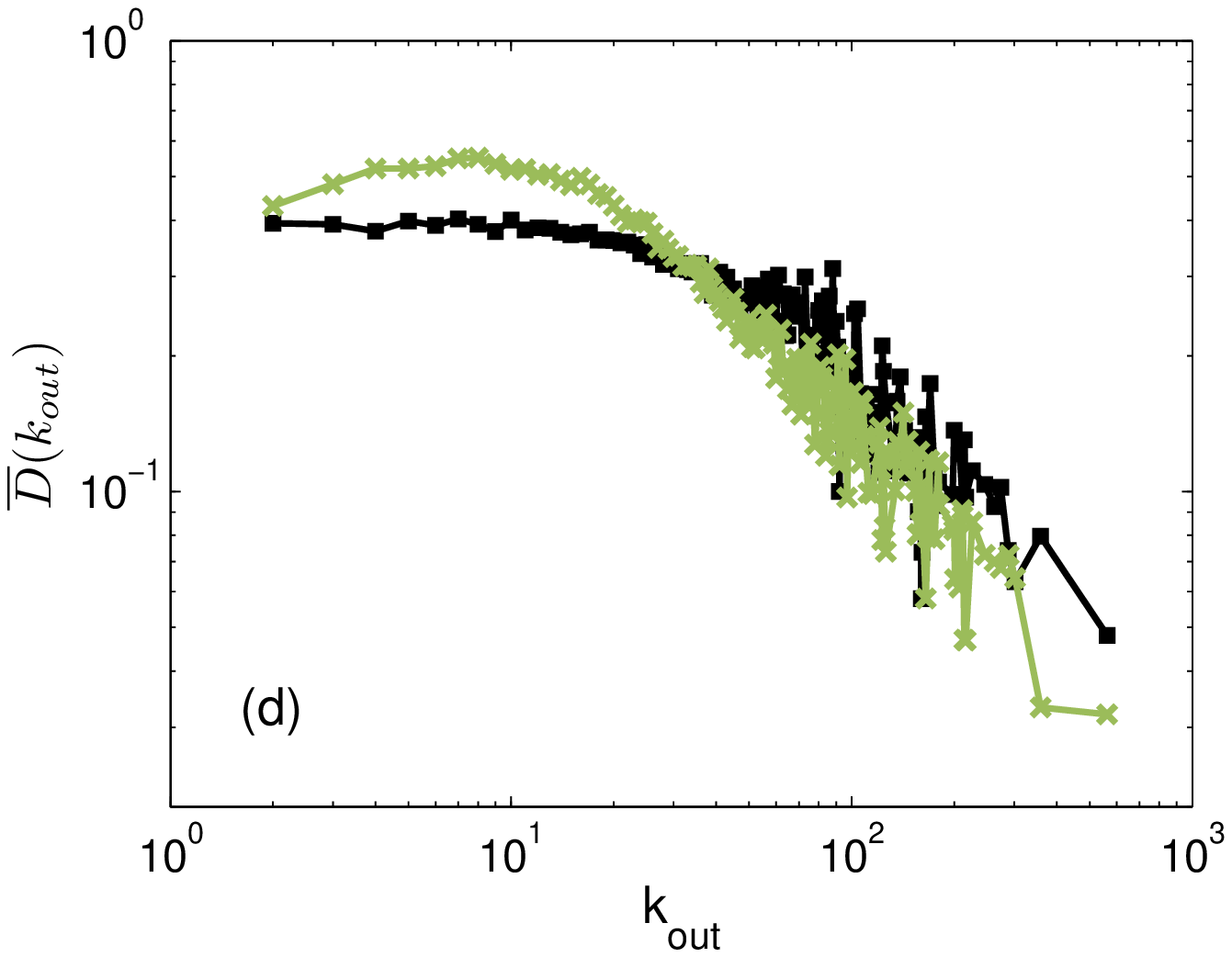}
\includegraphics[width=5.5cm]{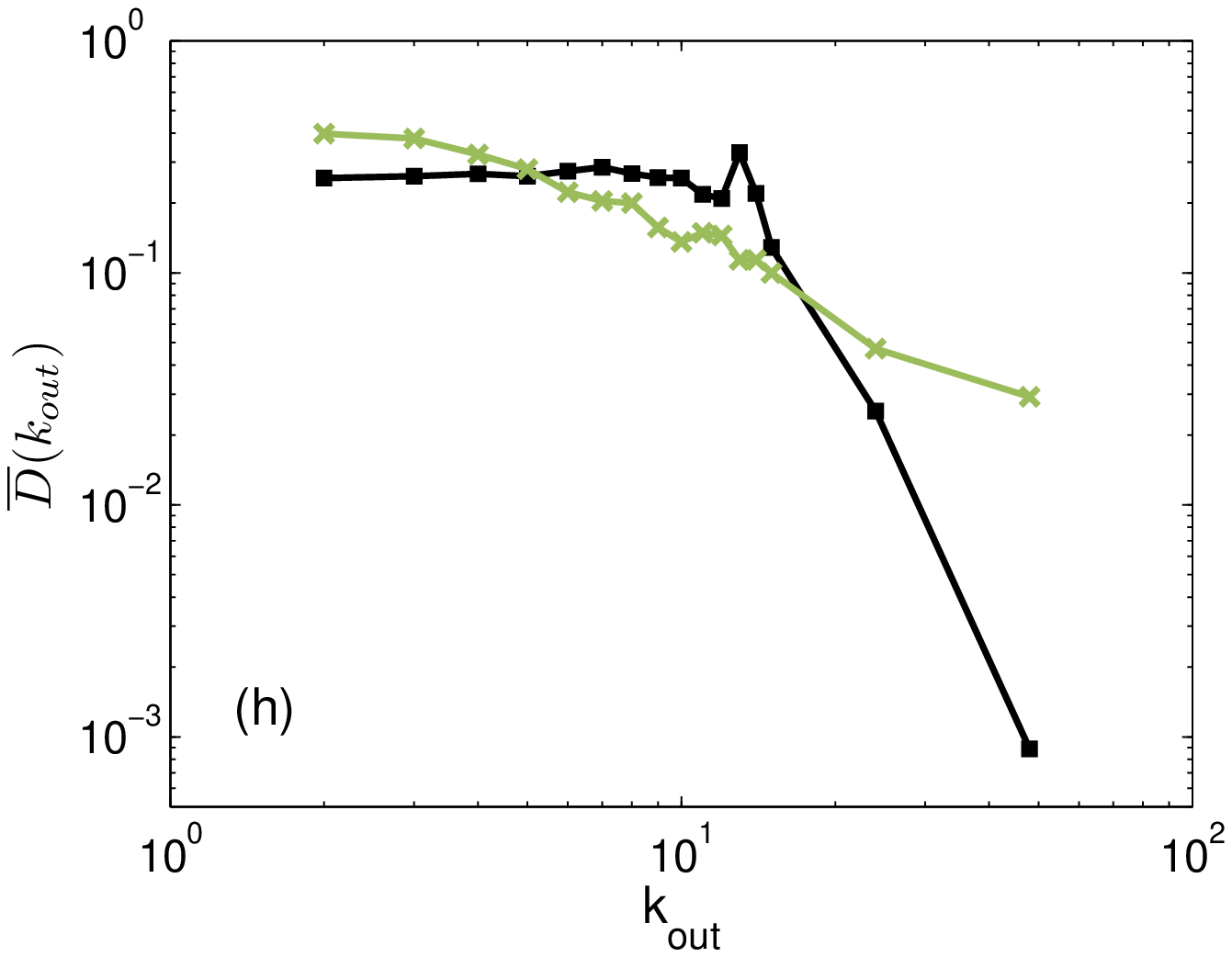}
\caption{\label{fig4} The in-degree distribution,
growth of triangle number $T_i$ as a function of network size $i$,
the average clustering coefficient $\overline{C}$ as a function of
node's degree $k$ and the link density of reference graph $\overline{D}$
as a function of node's out-degree $k_{out}$ of the two empirical networks and the DAC model.
Plots (a), (b), (c) and (d) are for hep-th data and (e), (f), (g) and (h) are for PNAS data.
Parameters in the model are scanned in their reasonable ranges and gained by
the best fit for the empirical data, i.e., $\alpha=1$ and $\beta=0.48$ for hep-th data
and $\alpha=1$ and $\beta=0.44$ for PNAS data. The results are averaged
over 100 independent realizations.}

\end{figure}

The numerical results are shown in table~\ref{tab:Data_description}
and Fig.~\ref{fig4}.
We find that although the two data are very different in nature,
many structural characteristics are shared,
i.e., the in-degree distributions both follow a power law,
the triangle numbers are both much larger than random
networks and the number of triangles both follow a
similar growth law as a function of the network size.
For the performance of our DAC model, in table~\ref{tab:Data_description}
we see the number of triangles and the average clustering coefficient
are both matched for the two data, which confirms that our model
can reproduce the clustering features of citation networks.
Detailed comparisons are shown in Fig.~\ref{fig4}.
For the hep-th data, as Fig.~\ref{fig4}(a) shows,
the in-degree distribution is well fitted.
In Fig.~\ref{fig4}(b), we can see that not only the final number,
but also the growth of the triangle number is remarkably matched
between our model and the empirical data. Fig.~\ref{fig4}(c) reveals
that the average clustering coefficient decays with the node's degree in the
data and this feature is captured by our DAC model.
The fourth quantity is the link density of reference graph that we show
in Fig.~\ref{fig4}(d). The relationship between link density and
out-degree is well reproduced by the DAC model.
For the PNAS data, the four statistics observed here are all
well reproduced by the model too.

\begin{figure}
\centering
\includegraphics[width=5.5cm]{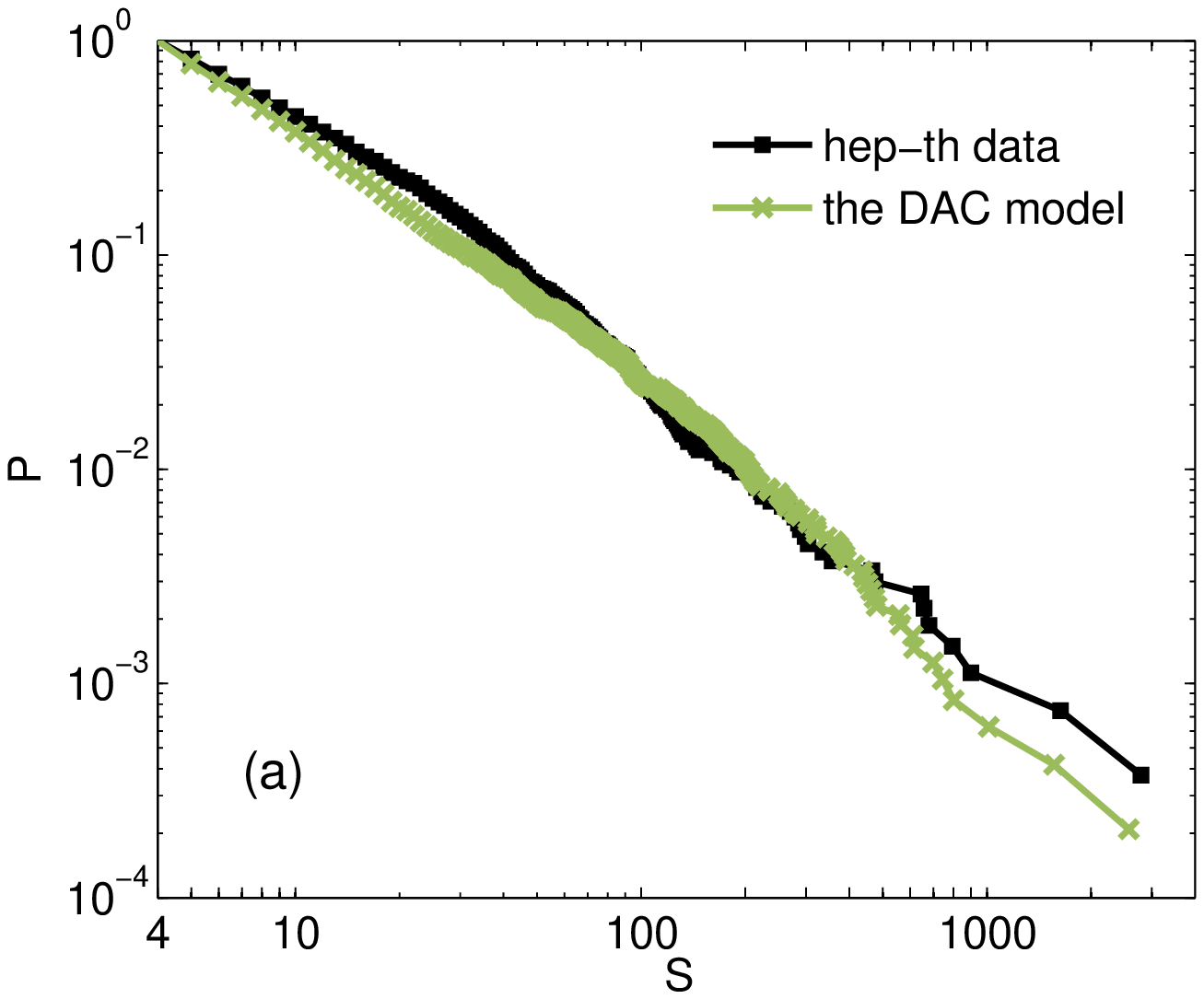}
\includegraphics[width=5.5cm]{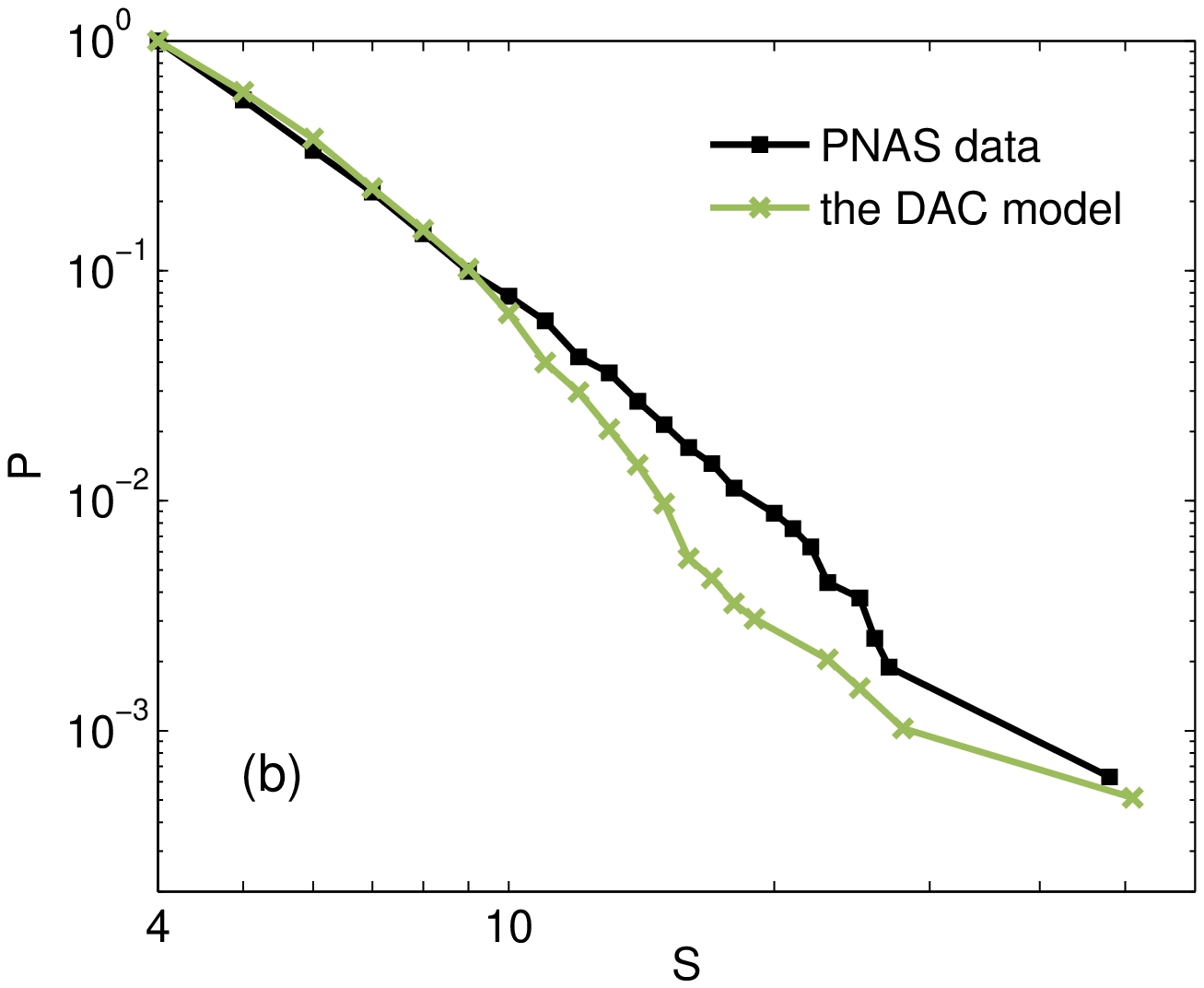}
\caption{\label{fig5} The cumulative size distribution of co-citation clusters
for the two empirical networks and the DAC model. The clusters are gained by 
clique percolation method \cite{Palla2005} with $k=4$. In both figures, $P$ denotes 
the cumulative distribution and $S$ denotes the size of clusters.}

\end{figure}

Besides the microscopic clustering statistics such 
as number of triangles and clustering coefficient, 
we also investigate the size distribution of co-citation clusters \cite{Chen2010} 
to verify the effectiveness of our model. For a given citation network, 
we first construct a co-citation network, in which nodes are papers and two nodes 
add one link once their corresponding papers are cited by a same paper. 
The co-citation network is undirected and weighted with weight 
on edge $e_{ij}$ measured in terms of cosine coefficients between 
the two sets of papers that cite i and j respectively \cite{Chen2010}.
Then we use the clique percolation method (CPM) \cite{Palla2005} to 
identify co-citation clusters in the co-citation network. 
As CPM requires the network to be unweighted, we remove all edges 
with weights smaller than a threshold $w^*$ and $w^*$ is determined 
using the method provided in \cite{Palla2005}. 
Fig.~\ref{fig5} shows the size distributions of co-citation 
clusters for hep-th network and PNAS network. We can see that 
the DAC model generates comparable size distributions as the
real data. Moreover, the size distributions of the two networks both have
broad ranges, which is in agreement with the results in \cite{Palla2005}. 

\section{Conclusion and Discussion}
\label{sec5}

In this paper, we focused on modeling the clustering features in citation networks.
We observed that the reference graphs are always highly connected and
contain lots of cliques, which helps the formation of clustering in the network.
Based on these observations, we proposed a growth model, the DAC model, for citation networks.
The model adds nodes one by one and fills up the nodes' out-degrees taking
advantage of two attachment mechanisms: the degree-aging preferential attachment
and the clique neighbourhood attachment. We validated the model by comparing
four quantities, the in-degree distribution, the growth of triangle number,
the average clustering coefficient, the link density of reference graphs and the size 
distribution of co-citation clusters on two real-world citation 
networks. Good agreements are gained for both data
by tuning parameters in the attachment mechanisms.

The results on the two real-world data suggest that the attachment mechanisms
in the model capture the linking rules of scientific citation networks:
a paper prefers to cite recent and popular ones and this
helps to form the degree distribution of the network.
Moreover, a paper tends to cite its neighbours' clique neighbours and this
helps to form the clustering. This work is a step forward in the modeling of citation
networks and will provide insights for further studies such as the formation
of subgraphs.

In this paper we provide one way to incorporate the topological information
of the potential neighbours and better methods are worth being explored.
Nodes in citation networks are always documents, thus
textual or semantical information may be helpful
in the preferential attachment mechanisms and
the previous works \cite{Menczer2004,Cheng2009} give us
some indications. Moreover, high clustering is a common characteristic in many
real-world networks and we will further test our mechanisms in modeling
the evolutions of other kinds of network, such as the social network.

\section*{Acknowledgements}
We acknowledge Zhi-Xi Wu for providing code and discussion about
their triadic closure model for citation networks. We also thanks Tao Zhou
for helpful suggestions. This work is partially
supported by the National Natural Science
Foundation of China under grant numbers 60873245, 60933005 and 61173008.

\end{document}